\newif\ifprp
 \prptrue   % Preprint
\ifprp
 \documentstyle[twoside,aps,prb,epsf,floats]{revtex} 
\else
 \documentstyle[aps,prb]{revtex}
\fi

\begin{document}

\ifprp
 \twocolumn[\hsize\textwidth\columnwidth\hsize\csname 
 @twocolumnfalse\endcsname 
\fi

\draft

\title{Spin  Splitting  and  Weak  Localization  in  (110) GaAs/AlGaAs
Quantum Wells}

\author{T. Hassenkam, S. Pedersen,  K. Baklanov, A. Kristensen,
C. B. Sorensen, and P. E. Lindelof}

\address{The   Niels   Bohr   Institute,   University  of  Copenhagen,
Universitetsparken 5, DK-2100 Copenhagen, DENMARK}

\author{F. G. Pikus$^\dagger$}

\address{University of California, Santa Barbara CA 93106, USA}

\author{G. E. Pikus}

\address{A.F. Ioffe Physicotechnical  Institute 194021 St  Petersburg,
RUSSIA}

\date{\today} 
\maketitle 

\begin{abstract}
We investigate experimentally and theoretically the spin-orbit effects
on the weak  localization in a  (110) GaAs 2-dimensional  electron gas
(2DEG).  We  analyze  the  role  of  two  different  terms in the spin
splitting of the conduction  band: the Dresselhaus terms,  which arise
due to the lack of inversion  center in the bulk GaAs, and  the Rashba
terms, which are caused  by the asymmetry of  the quantum well. It  is
shown that in $\rm A_3 B_5$ quantum wells the magnetoresistance due to
the weak localization depends qualitatively on the orientation of  the
well. In particular, it is demonstrated that the (110) geometry has  a
distinctive  feature  that  in  the  absence  of  the Rashba terms the
``antilocalization'' effect, i.e. the positive magnetoresistance, does
not exist. Calculation of the weak anti-localization magnetoresistance
is found to be in excellent agreement with experiments.
\end{abstract}

\ifprp
 \vskip 2pc ] % end \twocolumn[...] 
\fi


\narrowtext

\section{Introduction}

The effect of the negative magnetoresistance observed in  high-density
2$d$ electron gas in semiconductor quantum wells is known to be caused
by  the  weak  localization,  which  results  from  the   constructive
interference of two electron waves propagating along a closed path  in
opposite directions, and leads to suppression of the conductivity.  In
a  magnetic  field  the  interference  conditions  are violated, which
causes   the   effect   of   the   {\em   negative  magnetoresistance}
\cite{Altshuler80}.

It was shown in \cite{Hikami80} that the triplet states with the total
momentum  of  both  electron  wavefunctions  $J=1$  give  a   positive
contribution into the resistance,  while the singlet state  with $J=0$
gives   a   negative   contribution   (antilocalization).   Then,  the
interference conditions can  also be changed  by the spin  relaxation,
which,  depending  on  the  relaxation  mechanism,  can  suppress  the
contribution       of       either       triplet or (mainly) singlet       
states\cite{Hikami80,Altshuler81}.    In    the    non-centrosymmetric
semiconductors  and   semiconductor  structures   the  dominant   spin
relaxation mechanism is the Dyakonov--Perel mechanism, which is caused
by     the     spin     splitting     of     the    conduction    band
\cite{Dyakonov71,Pikus84}. If this splitting is not very small,  in weak magnetic fields  the
antilocalization effect  prevails, and  the resistance  increases with
the magnetic field $B$. Therefore, the nature and strength of the spin
relaxation  determines  not  just   the  magnitude  of  the   negative
magnetoresistance effect,  but even  the qualitative  behavior of  the
magnetoconductivity $\sigma(B)$.  Furthermore, it  was recently  shown
\cite{Iordanskii94,Pikus95,Knap96} that  if the  conduction band  spin
splitting is linear in  the wave vector, which  is always the case  in
2$d$ structures, the  theory of the  weak localization must  take into
account the correlation between the electron motion in co-ordinate and
spin   spaces.   The   effects   of   the   spin   relaxation  on  the
magnetoresistance were  recently investigated  experimentally by  Knap
{\it  et   al.\/}\cite{Knap96}  and   by  Pedersen   {\it  et   al.\/}
\cite{Pedersen96}  and  a  very  good  agreement was obtained with the
theory for  the (100)  oriented GaAs  quantum wells.  In this paper we
report  for  the  first  time  on  a  study  of  magnetoresistance  in
(110)-oriented  quantum  wells  and  present  the  theory  of the weak
localization for this particular case.

\section{Theory}

In asymmetric ${\rm A}_3 {\rm  B}_5$ quantum wells the spin  splitting
of  the  conduction  band  has  two  terms.  The  first,   Dresselhaus
term\cite{Dresselhaus55},  arises  from  the  asymmetry of the crystal
itself  and  in  the  bulk  crystal  is  described  by  the  following
Hamiltonian:

\begin{equation}
{\cal H}_1 = \gamma \sum \sigma_i k_i \left(k_{i+1}^2 - k_{i+2}^2\right),
\label{HamBulk}
\end{equation}

\noindent where $i = x, y, z$, $i+3 \rightarrow i$,
$\gamma$  is the spin-orbit  coefficient for the  bulk
semiconductor,  $\sigma_i$  are  the  Pauli  matrices  and  $\vec{k}$ is the
electron wave  vector (in  this paper  we take  $\hbar = 1$ everywhere
except in final formulas). We take the coordinate system $z \parallel 110$,
$x  \parallel  1\bar{1}0$,  and  $y  \parallel  001$.
In a (110) quantum well the $k_z$ is quantized: $\langle k_z
\rangle  =  0$,  $\langle  k_z^2  \rangle  =  \int  \left| \nabla \psi
\right|^2 \, dz$, where $\psi(z)$ is the electron wave function in the
well. Consequently,
  the Hamiltonian (\ref{HamBulk}) becomes

\begin{equation}
{\cal H}_1 = - \gamma \sigma_z k_x \left[ \frac{1}{2} \langle k_z^2 \rangle -
\frac{1}{2} \left( k_x^2 - 2 k_y^2 \right) \right].
\label{HamDr}
\end{equation}

\noindent  It  is  convenient  to  write  this Hamiltonian as a sum of
harmonics\cite{Iordanskii94,Pikus95,Knap96}:

\begin{equation}
{\cal H}_1 = \sigma_z \left( \Omega_{1z} + \Omega_{3z} \right),
\label{HamDrHarm}
\end{equation}

\noindent where

\begin{eqnarray}
\Omega_{1z} & = & \Omega_1 \cos \phi, \
\Omega_{3z} = \Omega_3 \cos 3\phi, \
\nonumber \\
\Omega_1 & = & -\frac{1}{2} \gamma k \left(\langle k_z^2 \rangle -
\frac{1}{4} k^2 \right), \
\Omega_3 = \frac{3}{8} \gamma k^3,
\label{Omega} \\
k^2 & = & k_x^2 + k_y^2, \ \tan \phi = \frac{k_y}{k_x}.
\nonumber
\end{eqnarray}

The other term in the conduction band spin splitting, the Rashba term,
is caused  by the  asymmetry of  the quantum  well\cite{Rashba61}. Its
Hamiltonian {\em  does not  depend on  the orientation  of the quantum
well}:

\begin{equation}
%{\cal H}_2 = \left[\vec{\sigma} \times
%\vec{\Omega}_2\right]_z,
{\cal H}_2 = \left(\vec{\sigma} \cdot 
\vec{\Omega}_2\right),
\label{HamRash}
\end{equation}

\noindent where  $\Omega_{2_x} =  \Omega_2 \sin\phi,  \Omega_{2_y} = -\Omega_2
\cos\phi$, $\Omega_2 = \alpha k$. In a uniform electric field $\cal E$
(triangular  well)  $\alpha  =  \alpha_0  e  \cal  E$; the coefficient
$\alpha_0$ may depend on the properties of the heterointerface.

Using the formalism similar to that of Refs.~\cite{Iordanskii94,Pikus95,Knap96} one can show that
the  correction  to  the  conductivity  $\sigma$  caused  by  the weak
localization is determined by the zero harmonic of the Cooperon ${{\rm
\kern.24em  \vrule  width.05em  height1.4ex  depth-.05ex   \kern-.26em
C}}_0(\vec{q})$, which obeys the following equation:

\begin{equation}
{\cal H}
{{\rm \kern.24em
            \vrule width.05em height1.4ex depth-.05ex
            \kern-.26em C}}_0
= {1 \over 2 \pi \nu_0 \tau_0^2},
\label{Cooperon}
\end{equation}

\noindent where $\nu_0$ is the  density of states at the  Fermi level,
$\tau_0$ is the elastic lifetime and $v$ is the Fermi velocity. In the
basis of the eigenfunctions $\phi_0$ (antisymmetric singlet state) and
$\phi_l^m$ with $l = 1$, $m = -1, 0, 1$ (symmetric triplet state)  the
operator  ${\cal  H}$  consists  of  two  blocks, ${\cal H}_0$ for the
singlet states and $\tilde{\cal H}$ for the triplet states:

\begin{eqnarray}
{\cal H}_0 & = & D (q_x^2 + q_y^2) + {1 \over \tau_\varphi}, \
\tilde{\cal H} = D (q_x^2 + q_y^2) + {1 \over \tau_\varphi}
\label{Er} \\
& & + 2 \bigg[2 \Omega_2^2
+ J_z^2 \left(\Omega_1^2 - \Omega_2^2 +
\Omega_3^2 \frac{\tau_3}{\tau_1}\right) - 2 J_y J_z \Omega_1
\Omega_2\bigg] \tau_1
\nonumber \\
&& + 2 v \tau_1 \bigg[q_x \left(\Omega_1 J_z - \Omega_2 J_y\right) +
q_y \Omega_2 J_x \bigg].
\nonumber
\end{eqnarray}

\noindent Here $J_i$ are the matrices of the angular momentum operator
with  total  momentum  $J  =  1$, $\tau_\varphi$ is the phase-breaking
time, $D = v^2 \tau_1/2$ is the diffusion coefficient, $\tau_n$, $n  =
1,  3$,  is  the  relaxation  time  of the respective component of the
distribution function.

In a magnetic field $B \parallel z$ the wave vector $\bf q$ becomes an
operator with the commutation relations

\begin{equation}
[q_+q_-] = {\delta \over D}, \ \delta = {4 e B D \over \hbar c},
\label{Commut}
\end{equation}

\noindent where $q_\pm = q_x \pm i q_y$.

\noindent  This  allows  us  to  introduce  creation  and annihilation
operators $a^\dagger$ and $a$, respectively, for which $[aa^\dagger] =
1$:

\begin{equation}
D^{1/2} q_+ = \delta^{1/2} a,
\quad
D^{1/2} q_- = \delta^{1/2} a^\dagger,
\quad
D q^2 = \delta \{a a^\dagger\}.
\label{OperQ}
\end{equation}

\noindent The weak  localization correction to  the conductivity in  a
magnetic field can now be written as

\begin{equation}
\Delta \sigma = - {e^2 \delta \over 4 \pi^2 \hbar} \sum_{n=0}^{n_{max}}
\left(- {1 \over {\cal E}_{0n}} + \sum_{m=-1}^1 {1 \over E_{mn}}\right),
\label{NMRB}
\end{equation}

\noindent  where  $n_{max}  =  1/\delta\tau_1$. The eigenvalues ${\cal
E}_{0n}$ of ${\cal H}_0$  are given by the following equation:

\begin{equation}
{\cal E}_{0n} = \delta \left( n + \frac{1}{2} \right) +
\frac{1}{\tau_\varphi}.
\label{E0}
\end{equation}

\noindent  The expression
for  the  operator  $\tilde{\cal  H}$,  of  which  $E_{mn}$  are   the
eigenvalues, follows from Eqs.~(\ref{Er}, \ref{OperQ}):

\begin{eqnarray}
\tilde{\cal H} & = & \delta \{a a^\dagger\} + {1 \over \tau_\varphi} +
2 \left(\Omega_1^2 \tau_1 + \Omega_3^2 \tau_3\right) J_z^2
\nonumber \\
& & + 2 \left(2 - J_z^2\right) \Omega_2^2 \tau_1 - 4 \Omega_1 \Omega_2
\tau_1 J_y J_z
\label{ErMag} \\
& & + 2 (\delta \tau_1)^{1/2} \left[ \frac{1}{\sqrt{2}} \Omega_1 J_z
\left(a^\dagger + a\right) + i \Omega_2 \left(a^\dagger J_+ - a
J_-\right)\right].
\nonumber
\end{eqnarray}

\noindent where $J_\pm = (J_x \pm iJ_y)/\sqrt{2}$.

If we keep only the  Dresselhaus terms in Eq.~(\ref{Er}), i.e.  put
$\Omega_2 = 0$,  the matrix $\tilde{\cal  H}$ becomes diagonal  in the
basis of the eigenfunctions of $J_z$,  and its non-zero matrix elements
for arbitrary $n$ and $m = -1, 0, 1$ can be written as

\begin{equation}
\tilde{\cal H}_{mm} = D \left[ q_y^2 + \left(q_x + q_m\right)^2
\right] + 2 \Omega_3^2 \tau_3 m^2 + \frac{1}{\tau_\varphi},
\end{equation}

\noindent where $q_m = (2\Omega_1/v)m$. Since the shift by $q_m$  does
not change the commutation relations (\ref{Commut}) for the  operators
$q_y$ and $q^\prime_x \equiv q_x + q_m$, the energies $E_{mn}$  depend
only on the cubic Dresselhaus term:

\begin{equation}
E_{mn} = \delta \left(n + \frac{1}{2}\right) + 2 \Omega_3^2 \tau_3 m^2 +
\frac{1}{\tau_\varphi},
\label{EDres}
\end{equation}

\noindent while the spin relaxation  rate is determined by the  sum of
all terms: 
$\tau_s^{-1} = 2 \left(\Omega_1^2 \tau_1 + \Omega_3^2
\tau_3\right).
$
One  can see  from Eqs.~(\ref{NMRB},  \ref{E0}, \ref{EDres})
that the term with $m = 0$ cancels the contribution of ${\cal E}_{0n}$
in   the   conductivity,   and,   therefore,  the  magnetoconductivity
$\Delta\sigma(B)$ is given by the expression\cite{Hikami80}:

\begin{eqnarray}
& & \Delta \sigma(B) - \Delta \sigma(0) =
\nonumber \\
& & \ \
\frac{e^2}{2\pi^2 \hbar}
\left\{ \Psi \left( \frac{1}{2} + \frac{H_\varphi}{B} + \frac{H_{\rm
SO}^{(3)}}{B} \right) - \ln{H_\varphi \over B}\right\},
\end{eqnarray}

\noindent where

\begin{equation}
H_\varphi = {c  \hbar \over 4 e D  \tau_\varphi}{\rm ,} \ \
H_{\rm SO}^{(3)} = {c \hbar \over 4 e D} 2\Omega_3^2\tau_3.
\label{HPhi}
\end{equation}

\noindent 
Therefore, {\em in absence of the Rashba terms in a (110) quantum well
the negative magnetoresistance cannot be observed}.

When both Dresselhaus and Rashba terms are present, the eigenvalues of
$\tilde{\cal H}$ can be found only numerically. In practice it is more
convenient to  compute directly  the sum  of the  inverse eigenvalues,
using the expression\cite{Knap96,Pikus96}:

\begin{equation}
\sum_n \sum_{m = -1}^1 \frac{1}{E_{nm}} =
\sum_i \frac{\left|D_{ii}\right|}{|D|},
\label{Minors}
\end{equation}

\noindent where $|D|$ is the determinant of the matrix $\tilde{\cal H}$
and $|D_{ii}|$ is the minor of its diagonal element $i, i$. 
The detailed description of the numerical procedure will  be
published elsewhere.

\section{Experiment}

The samples  used in  our work  were grown  by Molecular  Beam Epitaxy
(MBE) technique. The layer sequence was of the standard high  mobility
transistor  type.  The  2$d$  electron  gas  was formed in GaAs at the
(110) $\rm
GaAs/Ga_{0.7} Al_{0.3}  As$ interface.  The sample  was $\delta$-doped
with silicon in two  planes at 10 nm  and 50 nm from  the interface 
The individual  samples were  mesa-etched into rectangular
Hall bars with the width of 0.2  mm and the total length of 4.2  mm. 3
voltage contacts on each side were  placed at a distance of 0.8  mm to
avoid  perturbing  significantly   the  4-point  measurements.   Ohmic
contacts to the 2DEG was  made by an annealed AuGeNiAu  composite film
in $0.6 \times  0.6 {\ \rm  mm^2}$ contact areas.  The contacted areas
were subsequently bonded to the legs of a non-magnetic chip carrier.

Our  4-point  measurements  of  the  resistivity  were  carried out by
standard  low-frequency  lock-in   technique.  Typically  the   sample
resistance was few  k$\Omega$ and with  an AC current  amplitude below
200 nA we have avoided significant Joule heating of the sample at  our
lowest temperature, 0.3  K. 
To generate low stable
magnetic fields we  used two highly  stable current sources  (Keithley
220); the first was used to out-compensate the magnetic flux trapped in
the  superconducting  magnet,  whereas  the  second  was  used for the
magnetic  field  sweep  around  the  zero  value. The peak in the weak
localization  resistivity  (or,  for  antilocalization,   conductance)
defines the zero value.  Incidentally, this method is  accurate enough
to  determine  this  zero  point   within  about  $1  \  \rm   \mu  T$
\cite{Rasmussen96}.

\ifprp
 \begin{figure}[t]
 \epsfxsize=3 in
 \epsffile{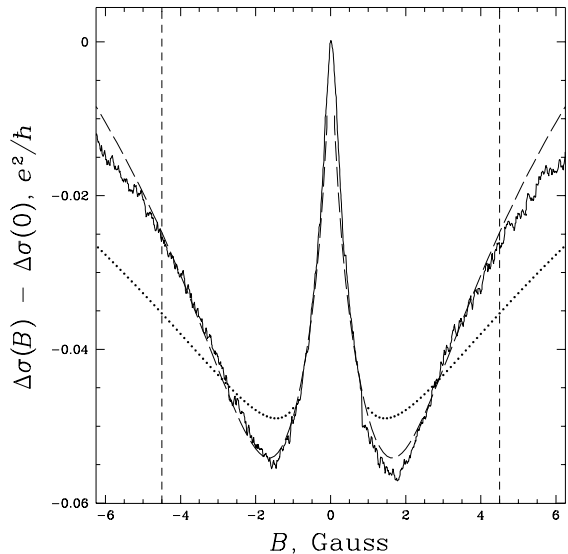}
 \caption{Magnetoconductivity  $\Delta  \sigma(B) - \Delta\sigma(0)$
 in (110) quantum well. Experimental results
 are shown by solid line, theoretical best fit -- by dashed line
 The dots show the best fit by the Hikami-Larkin-Nagaoka theory.
 Sample characteristics and parameters of the
 theory are given in the text.  The vertical
 lines show the interval $|B| \le H_{\rm tr} = 4.5 {\ \rm Gs}$.
 \label{Results}}
 \end{figure}
\fi

\section{Results and Discussion}

In Fig.~\ref{Results} we show  the results of the  magnetoconductivity
measurements for a sample with  electron density $n = 5  \cdot 10^{11}
{\ \rm cm^{-2}}$ and mobility $\mu = 7 \cdot 10^4 {\ \rm cm^2/V s}$ at
$T = 0.36 {\  \rm K}$. Also shown  are the best fits  as obtained from
our  theory,  and  from  the  theory  of  Hikami,  Larkin, and Nagaoka
(HLN)\cite{Hikami80},  which  assumes  that  all  terms  of  the  spin
splitting give  additive contributions  into magnetoconductivity.  The
fitting was done by  weighted explicit orthogonal distance  regression
using  the  software  package  ODRPACK  \cite{ODR}.  The  weights were
selected to  increase the  importance of  the low-field  ($B \le 3 \rm
Gs$)  part  of  the  magnetoconductivity  curve. Only the experimental
points at $|B| \le H_{\rm tr} = {c \hbar \over 4 e D \tau_1} = 4.5 \rm
Gs$ we used  for fitting, since  the above theories  use the diffusion
approximation, and, therefore, are  only valid for $B$  small compared
to $H_{\rm tr}$.

The parameters of our theory are $\tau_\varphi$ and $\Omega_i$,  $i=1,
2, 3$. It is convenient  to convert them into characteristic  magnetic
fields  $H_\varphi$, $H_{\rm SO}^{(3)}$ {\bf  (} see Eq.~(\ref{HPhi}){\bf )}, 
and   $H_{\rm SO}^{(1, 2)}$:

\begin{equation}
H_{\rm SO}^{(1)} = {c \hbar \over 4 e D} 2\Omega_1^2\tau_1, \
H_{\rm SO}^{(2)} = {c \hbar \over 4 e D} 2\Omega_2^2\tau_1.
\label{HSO} 
\end{equation}

\noindent The parameters of the best fit are $H_\varphi = 0.02 {\  \rm
Gs}$, $H_{\rm SO}^{(1)} = 0.12 {\ \rm Gs}$, $H_{\rm SO}^{(2)} = 1.3 {\
\rm Gs}$, and $H_{\rm SO}^{(3)} = 0.04 {\ \rm Gs}$ for our theory  and
$H_\varphi = 0.014 {\ \rm Gs}$, $H_{\rm SO} = 0.33 {\ \rm Gs}$ for the
HLN theory.  One can  clearly see  that the  HLN theory  is unable  to
describe  the   experimental  data.   The  disagreement   between  the
experiment and the HLN theory in our case is much more severe than for
(001) quantum  wells \cite{Pikus95,Knap96},  since the  effects of the
correlations  between  the  electron  motion  in  co-ordinate and spin
spaces is  much stronger  here: as  we have  shown above,  in a  (110)
quantum  well  the  linear  Dresselhaus  terms  have  no effect on the
magnetoconductivity (in the absence of the Rashba term), whereas in  a
(001) well  such a  dramatic cancellation  is only  possible when both
Rashba and Dresselhaus terms exist and are nearly equal.

 From the  above values  of the  parameters $H_{\rm  SO}^{(i)}$ we  can
determine the values of  the constants $\gamma$ and  $\alpha_0$, using
Eqs.~(\ref{Omega}, \ref{HPhi},  \ref{HSO}), $k_{\rm  F} =  \sqrt{2 \pi
N_{\rm  s}}$,  and  the   following  expressions  for  $\langle   k_z^2
\rangle$\cite{Ando82}  and  $\alpha$,  which  are  obtained  using the
standard    variational    wavefunction    for    electrons   at   the
heterointerface\cite{Fang66}:

\begin{eqnarray}
\langle k_z^2 \rangle = \frac{1}{4}
 \left({16.5 \pi e^2 m N_{\rm s} \over \kappa \hbar^2}
\right)^{1/3}\!\!\!\!,
%\nonumber \\
\alpha = e \bar{\cal E} \alpha_0, \
\bar{\cal E} = \frac{2 \pi e N_{\rm s}}{\kappa},
\end{eqnarray}

\noindent where $N_{\rm s}$ is  the electron density, $\kappa$ is  the
dielectric  constant,  $m$  is   the  effective  electron  mass,   and
$\bar{\cal E}$  is the  average electric  field in  the well.  The
resulting values of the coefficients are $\gamma \approx 22 {\ \rm  eV
\AA^3}$ and $\alpha_0 \approx 14 {\ \rm \AA^2}$. The value of $\gamma$
is very  close to  the previously  reported values,  both measured and
calculated,\cite{Pikus84,Vogl83,Kobayashi82,Cardona94,Cardona95,Jusserand95,Pikus95,Knap96},
including those  measured in  weak localization  experiments in  (001)
quantum wells\cite{Pikus95,Knap96}. The  value of $\alpha_0$  was only
measured in the  latter experiments and  reported to be  about $7.2 {\
\rm  \AA^2}$.  However,  unlike  $\gamma$,  which is the bulk material
coefficient,     $\alpha_0$     may     contributions     from     the
interface\cite{Gerchikov92,PikusUnp}. Their magnitude is not  reliably
known; therefore, we do not view the discrepancy between our value  of
$\alpha_0$ and the one measured in (001) wells as alarming. Also, from
the  value  of   $H_{\rm  SO}^{(3)}$  we   can  determine  the   ratio
$\tau_3/\tau_1  \approx  1/8$.  This   ratio  can  vary  from   1  for
short-range scattering to $1/9$  for scattering on remote  impurities,
and the experiment shows that in our samples those are practically the
only  source  of  scattering.  Lastly,  we  can  determine  the  phase
relaxation time $\tau_\varphi \approx 6 \cdot 10^{-10} {\ \rm s}$.

\section{Conclusions}

In  conclusion,  we  have  presented  new experimental and theoretical
studies  of  magnetoconductivity  caused  by  the weak localization in
(110)  GaAs  quantum  wells.  It  is  demonstrated  that  if  the spin
splitting of the conduction  band is linear in  the wave vector it  is
necessary to take  into account the  correlation between the  electron
motion in co-ordinate and spin  spaces. This correlation leads to  the
special  feature  of  the  (110)  geometry:  in  a perfectly symmetric
quantum   well,   when   the   Rashba   terms  are  absent,  the  weak
antilocalization effect,  which leads  to positive  magnetoresistance,
does not exist. The presence of the positive magnetoresistance in  our
samples is  a clear  signature of  the Rashba  terms in the conduction
band spin splitting. Our new theory achieves a good agreement with the
experiment  and  gives  the  values  for  the  parameters  of the spin
splitting which are in  agreement with previous optical  and transport
experiments and theoretical calculations.

We thank  W. Knap  communicating preprints  prior to  publication. The
experimental part of  our research was  carried out at  NBIfAFGs III-V
NANOLAB and supported by  the Center for Nanostructures  (CNAST) under
the MUP II program. F. G. P. acknowledges support by the NSF Grant DMR
93-08011, the  Center for  Quantized Electronic  Structures (QUEST) of
UCSB  and  by  the  Quantum  Institute  of UCSB. G. E. P. acknowledges
support by RFFI Grant 96-02-17849 and by the Volkswagen Foundation.

\ifprp
\else
 \begin{figure}
 \caption{Magnetoconductivity  $\Delta  \sigma(B) - \Delta\sigma(0)$
 in (110) quantum well. Experimental results
 are shown by solid line, theoretical best fit -- by dashed line
 The dots show the best fit by the Hikami-Larkin-Nagaoka theory.
 Sample characteristics and parameters of the
 theory are given in the text.  The vertical
 lines show the interval $|B| \le H_{\rm tr} = 4.5 {\ \rm Gs}$.
 \label{Results}}
 \end{figure}
\fi

\end{document}